\documentclass[prd,nofootinbib,twocolumn,showpacs]{revtex4}
\usepackage{graphics}
\usepackage{bm}

\begin{document}

\title{
Band structure in classical field theory
}


\author{Michael Salem and Tanmay Vachaspati}
\affiliation{Department of Physics, Case Western Reserve University,
10900 Euclid Avenue, Cleveland, OH 44106-7079, USA.}

\begin{abstract}
Stability and instability bands in classical mechanics are
well-studied in connection with systems such as described
by the Mathieu equation. We examine whether such band
structure can arise in classical field theory in the context
of an embedded kink in 1+1 dimensions. The static embedded kink 
is unstable to perturbations but we show that if the kink is 
dynamic it can exhibit stability in certain parameter bands. 
Our results are relevant for estimating the lifetimes of various 
embedded defects and, in particular, loops of electroweak Z-string.
\end{abstract}


\

\maketitle

\section{Introduction}
\label{introduction}

A wide variety of classical static solutions in field theories have
been constructed and they can play crucial roles in particle physics,
condensed matter systems, and cosmology. Commonly the static solutions 
exist because of non-trivial topology of the field theory and in these 
cases the solution is known as a ``topological defect''. Alternately 
the solution may exist in the form of a topological defect embedded in 
a theory where the requisite topology is absent. In these cases the 
static embedded topological defect is usually unstable and can decay.
We are interested in determining if this instability is also present
in embedded defects that are not static.

It is helpful to consider a mechanical analog of the unstable defect
solution. Consider a particle in two spatial dimensions with position
${\bf x}(t) = (x(t),y(t))$ and Lagrangian
\begin{equation}
L = {1\over 2} m {\bf v}^2 - V({\bf x})
\label{particleL}
\end{equation}
where
\begin{equation}
V({\bf x}) = {{k_x}\over 2} x^2 - {{k_y}\over 2} y^2 + {c\over 2} x^2 y^2
\label{particleV}
\end{equation}
with $k_x$, $k_y$ and $c$ being positive parameters. The extremum of
the potential is at ${\bf x}=0$ and is a saddle point. The positive
eigenvalue at the saddle point is along the $x$ direction and the
negative eigenvalue is along the $y$ direction. The equations of motion 
are:
\begin{equation}
m {\ddot x} = - k_x x - c x y^2
\label{xeq}
\end{equation}
\begin{equation}
m {\ddot y} = + k_y y- c x^2 y
\label{yeq}
\end{equation}
where overdots denote derivatives with respect to $t$.
The static solution is $x=0=y$ and is clearly unstable to perturbations
in the $y$ direction. This is seen by taking $y$ to be a perturbation 
and then linearizing the equation for $y$:
\begin{equation} 
m {\ddot y} = + k_y {y}
\label{deltayx=0}
\end{equation}
This equation has exponentially growing solutions and hence the
particle is unstable to ``rolling off'' in the $y$-direction.

Next consider a dynamic solution of the system:
\begin{equation}
x(t) = A \cos (\omega t) \ , \ \ \ y(t)=0
\label{dynamicsoln}
\end{equation}
where $\omega = \sqrt{k_x /m}$. Is this solution stable? 

Once again we proceed by perturbing the $y$ equation, and linearizing
in the perturbations. This gives
\begin{equation}
m {\ddot y} = + k_y y - c A^2 \cos^2 (\omega t) y
\label{deltayxne0}
\end{equation}
where $y$ denotes the perturbation.
This is the Mathieu equation  (Sec. 5.2, \cite{MorFes53}). 

There exist solutions to the Mathieu equation of the form
\begin{equation}
y(t) = e^{i\nu t} P(t) \ ,
\label{Mateqsolnform}
\end{equation}
a result which is derived using Floquet's theorem. The value of 
$\nu$ depends on the parameters $k_y/k_x$ and $cA^2/k_x$ 
and can be found as described in Sec. 5.2 of Ref. \cite{MorFes53}. If 
$\nu$ is imaginary, the original solution is unstable; if 
$\nu$ is real, the original solution is stable.

\begin{figure}
\scalebox{0.80}
{\includegraphics{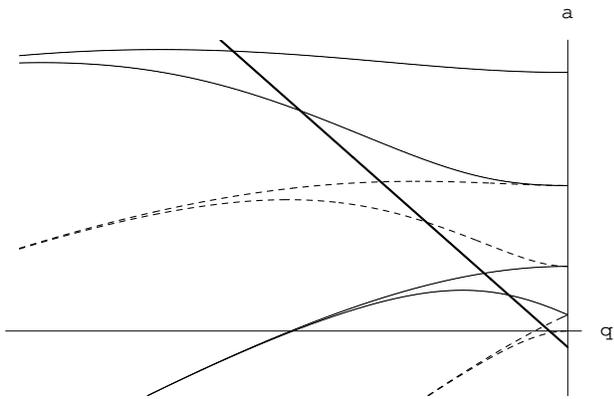}}
\caption{\label{Mathieueqbands}
Band structure for the Mathieu equation. For $q<0$, the
stability bands lie between consecutive dashed or consecutive
solid curves. The straight line denotes a set of parameters for 
the mechanical model for different values of the oscillation 
amplitude. Larger amplitudes correspond to points on the line 
that are further away from the origin. As the amplitude is increased, 
the line increasingly lies in the stability bands but no matter how 
large the amplitude, there are always bands where there is instability.}
\end{figure}

A remarkable feature of this system is that there are bands of stability
and instability. For fixed $\omega$, if $A$ is close to zero, there is 
an instability and the particle rolls off in the $y$-direction. For larger 
$A$, we find stability -- the particle gets displaced in the $y$-direction
at the saddle point, but is unable to roll off by the time it has moved
away from the saddle point in the $x$-direction. This is shown in
Fig. \ref{Mathieueqbands} where we have plotted the stability and
instability bands in the $a-q$ plane, where
\begin{equation}
a = {{cA^2}\over {2k_x}} - {{k_y} \over {k_x}} \ , \ \ 
q = - {{cA^2} \over {4k_x}} \ .
\label{qdefn}
\end{equation}
Then, for fixed $k_y/k_x$, we get the line
\begin{equation}
a = -2q - {{k_y} \over {k_x}}
\label{aqline}
\end{equation}
on the plot. This line intersects the unstable and stable regions
alternately as we go to larger amplitudes for fixed model parameters.

It is tempting to think that the pattern of stability and instability 
as the amplitude is increased can be understood in terms of the time 
spent by the particle in the vicinity of the saddle point as compared 
to the decay time of the static solution. However, this intuition does 
not work since for yet larger $A$ we again get instability. The band 
structure then has to be understood as a resonance phenomenon between 
the various oscillatory modes of the particle.

If we include frictional forces on the particle as it moves on
the potential, the amplitude of oscillations will gradually diminish
and eventually the particle will enter a band of instability. The
decay time of the oscillations will then be determined by the time
it takes for the particle to slip from a stability band to an
instability band.

We are interested in determining if a similar phenomenon can occur
in classical field theories. If a certain static field solution is
unstable, can its lifetime be much longer due to some dynamics? There
is clearly an affirmative though trivial answer -- since these are
relativistic field theories, a boosted field solution lives longer
due to relativistic time dilation. However, we are not interested in
this factor and would like to determine if there is an effect similar to
the band structure observed in the mechanical problem. There is another
complication in field theory: the dynamical solution will almost
certainly emit radiation and dissipate and hence is similar to the
mechanical case with the inclusion of friction. Yet in the examples
we will consider, the dissipation due to radiation is very small
and can be ignored.

We will restrict our attention in this paper to classical field
theory in 1+1 dimensions. Our first approach to the problem is
to construct unstable (embedded) kink solutions and study the growth
of linearized perturbations on kink solutions that are forced to
oscillate. The similarity between this field theory problem and
the mechanical problem described above can be understood by going
to the rest frame of the kink as it oscillates. In this (non-inertial)
frame the perturbation modes are oscillating in the vicinity of the 
kink. In the time when a perturbation mode lies outside the kink, 
it cannot grow. The mode can only grow during the time that it lies 
within the kink. This is exactly like the mechanical model where the 
particle (analogous to the perturbation mode) can roll off in the 
$y$-direction near the saddle point but not when its oscillations 
take it away from $x=0$. Hence we expect the behaviour of the 
perturbation mode to be similar to that of the particle.

Indeed our analysis reveals band structure in the growth of the
perturbation modes though we only find stability when the amplitude 
of the kink oscillations is comparable to the width of the kink.
In other words, in the linearized analysis only 
``jittering'' kink backgrounds are found to have stability bands. 
We have also performed a second analysis of the problem. 
In this approach, we impose an external potential in which the kink 
can oscillate. Now we evolve the full field equations and find the 
lifetime of the kink. Here again we find band structure. Furthermore, 
the amplitude of motion can be significantly larger than the kink width.
We attribute the enhanced stability bands at large amplitudes to the 
reduced and varying kink width in the external potential.
 
Our study has obvious implications for the study of classical
solutions that are unstable when static. Examples of such solutions
include a variety of embedded defects \cite{VacBar92, BarVacBuc94}. 
In particular, electroweak Z-strings are an important example 
\cite{Vac92, AchVac99}. Assuming, for now, that the band structure 
exists for dynamical solutions in 3+1 dimensions as well, we discuss 
the possible relevance of our work to the electroweak model in 
Sec. \ref{conclusions}.

\section{Linearized analysis of field theory problem}
\label{linearanalysis}

The 1+1 dimensional field theory we choose to study is one that 
contains embedded kink solutions \cite{VacBar92, BarVacBuc94}. 
The complex scalar field, $\phi = \phi_1 + i\phi_2$, has the 
Lagrangian density:
\begin{equation}
L =  {1\over 2}|\partial_\mu \phi |^2 - V(\phi ) 
\label{Lagrangian}
\end{equation}
where 
\begin{equation}
V(\phi ) = {1 \over 4} ( |\phi |^2 - 1 )^2 +
            {{(c-1)}\over 2} \phi_1^2 \phi_2^2 \ .
\label{potential}
\end{equation}
(In 1+1 dimensions, the fields are dimensionless. For convenience,
we have also rescaled the coordinates so that they are dimensionless.)
The last term in the potential is analogous to the cross-term in 
eq. (\ref{particleV})
and it violates the $U(1)$ symmetry $\phi \rightarrow exp(i\alpha )\phi$
when $c\ne 1$ but preserves a $Z_2 \times Z_2$ subgroup. 
A vacuum expectation value of either the real ($\phi_1$) or imaginary 
component ($\phi_2$) of $\phi$ spontaneously breaks the symmetry to 
the identity group.  Hence there are topological kink solutions in the 
model for $c\ne 1$. However, we will not be interested in these
topological solutions. Instead we will study the embedded kink solution 
which is not topological and exists for all values of $c$. This solution 
is:
\begin{equation}
\phi_1 = \tanh \left ( {x\over {\sqrt{2}}} \right ) \ , \ \ 
\phi_2 = 0
\label{embkinksoln}
\end{equation}
In the sub-space of field configurations defined by
$\phi_2 =0$, the embedded kink is topological and hence
stable. However, in the full field space, the static embedded
kink is unstable to perturbations in the $\phi_2$ direction.
The equation for linearized perturbations in the $\phi_2$
direction in the static embedded kink background is:
\begin{equation}
{\ddot \phi}_2 - {\phi}_2 '' + (c \phi_1^2 -1) \phi_2  = 0
\label{phi2perteq}
\end{equation}
where $\phi_1$ is given in eq. (\ref{embkinksoln}).
The solution for $\phi_2$ is a hypergeometric
function (Sec. 6.3, \cite{MorFes53}):
\begin{eqnarray}
\phi_2 & = &e^{i\nu t} {\rm sech}^K \left ({x\over \sqrt{2}} \right ) 
        \nonumber \\
        &&           F \left ( K+{1\over 2}+P,
                           K+{1\over 2}-P|K+1|z
                     \right )
\label{phi2solution}
\end{eqnarray}
with 
\begin{equation}
z = {1\over 2} \left [ 
      1 + \tanh \left ( {x\over \sqrt{2}} \right ) 
              \right ] \ ,
\end{equation}
\begin{equation}
K^2 = 2(c-1-\nu^2) \ ,
\end{equation} 
\begin{equation}
P= \sqrt{2c + {1\over 4}} \ .
\end{equation}
The lowest eigenvalue $\nu_0$ is given by:
\begin{equation}
\nu_0 ^2 = {1\over 2}\sqrt{2c+{1\over 4}} -{5\over 4} 
\label{nu}
\end{equation}
Since $\nu_0$ is imaginary for $1\le c < 3$, the static
solution in eq. (\ref{embkinksoln}) is unstable for 
these parameter values. For $c=3$, the lowest value of $\nu$
is zero, corresponding to the translation mode of the topological
$Z_2$ kink \cite{Raj87}. Hence for $c \ge 3$, the static
embedded kink is stable.

We now want to introduce some dynamics to the embedded kink solution 
and study the effect of the dynamics on the unstable mode in the $\phi_2$
direction. In the linearized approach to the problem, the dynamics is 
introduced ``by hand'' by introducing a time dependence in the unperturbed 
background:
\begin{equation}
\phi_1= 
    \tanh \left [ {\gamma \over \sqrt{2}} (x-g(t)) \right ] \ , 
\ \
\phi_2 = 0
\end{equation}
where 
\begin{equation}
g(t) = A \cos (\omega t)
\end{equation}
describes oscillations with amplitude $A$ and angular frequency
$\omega$, and $\gamma$ is the ``Lorentz factor''
\begin{equation}
\gamma = {1 \over {\sqrt{1-{\dot g}^2}}}
\label{gamma}
\end{equation}
We require $\omega A < 1$ so that the speed of the kink is
always sub-luminal.
In this background the equation for $\phi_2$ perturbations is:
\begin{equation}
{\ddot \phi}_2 - \phi_2 '' + (c \tanh^2 X -1) \phi_2 = 0
\label{linearphi2eq}
\end{equation}
where
\begin{equation}
X (t,x) \equiv {\gamma \over {\sqrt{2}}} (x-g(t)) \ .
\label{Xdefn}
\end{equation}
Note that boosts of the static solution can be obtained by setting
$g = v t$ where $v$ is the speed of the kink. Such backgrounds do
have a longer lifetime but this is due to the special relativistic
time dilation.

In Appendix A we extend Floquet's theorem to field theory 
to show that solutions of eq. (\ref{linearphi2eq}) have the form:
\begin{equation}
\phi_2 (t,x) = e^{i\nu t} P(t,x)
\label{phi2floquet}
\end{equation}
where $P(t,x)$ is a periodic function in $t$:
$$
P(t+2\pi \omega^{-1} , x) = P(t,x)
$$
The next step is to find the eigenvalue $\nu$. We were unable to
determine $\nu$ analytically and instead solved eq. (\ref{linearphi2eq})
numerically for various parameters $A$, $\omega$ and $c$ and with
an initial perturbation:
\begin{equation}
\phi_2 (t=0, x) = {\rm sech} (X(0,x))
\label{phi2t=0}
\end{equation}

\begin{figure}
\scalebox{0.80}
{\includegraphics{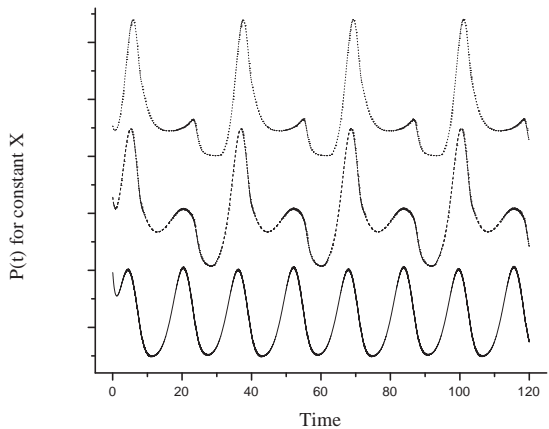}}
\caption{\label{numericalevoln}
The spatial value of the function $P(t,x)$ for 
$X = 0$ (solid curve), 1 (dashed curve), 2 (dotted curve) as 
a function of time. 
}
\end{figure}

\begin{figure}
\scalebox{0.80}
{\includegraphics{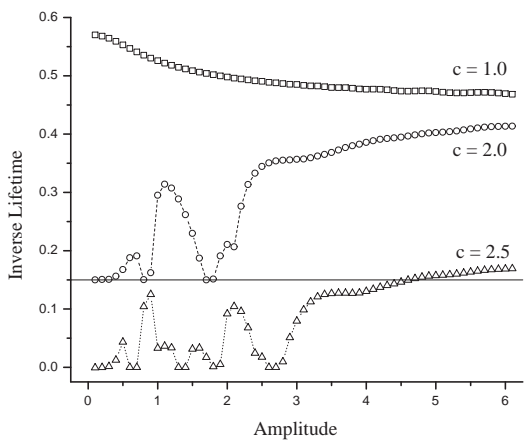}}
\caption{\label{tau-1vsA}
The decay rate $\tau^{-1}$ is plotted versus the amplitude of
oscillation for $c =$ 1 (solid curve), 2 (dashed curve), and 
2.5 (dotted curve), as determined in the linearized analysis. 
Stability bands are regions where $\tau^{-1} =0$. There are no 
stability bands for $c=1$ but these do exist for $c = 2$ and 2.5.
The curve for $c=2$ has been off-set by 0.15 along the y-axis for 
clarity.
}
\end{figure}

A sample result of the numerical evolution is shown in 
Fig. \ref{numericalevoln}. Here we plot $P(t,x)$ for several
different values of $X$ (see eq. (\ref{Xdefn})) versus time.
The plots are consistent with the form in eq. (\ref{phi2floquet}). For 
some values of $A$, $\phi_2 (t,x)$ grows exponentially with
time, while for others it oscillates without growth. This allows us 
to find the lifetime, $\tau$, of the background. In Fig. \ref{tau-1vsA}, 
we show the inverse lifetime, $\tau^{-1}$, as a function of the amplitude 
$A$ for $c=1, 2, 2.5$ with $\omega$ fixed by the relation $\omega = 0.99/A$. 
The reason we held $\omega = 0.99/A$ is that we expect that
stability will depend on the acceleration $a_c$ of the kink and this
is $\le \omega^2 A$. Since the speed cannot exceed the speed
of light, we impose $\omega A \le 1$. Therefore $a_c \le v_{max}^2/A$
where $v_{max} = \omega A$ is the maximum speed of the kink.
So we expect greatest stability when $v_{max} \sim 1$ and for small 
oscillation amplitude. In Fig. \ref{numericalevoln}, the bands
of stability occur where the inverse lifetime vanishes.

It is worth noting that the values of $A$ for which stability can occur 
are of order the half-width of the kink which is given by $\sqrt{2}$. 
Hence, stability bands only occur when the background is 
undergoing a rigid jittering motion. The point $A=0$ is a singular
point in our approach since the oscillation frequency is
$\omega = 0.99/A$ and this becomes infinite as $A$ tends to
zero. Another important point is that 
there are no stability bands for the embedded kink in the $U(1)$ model 
obtained with $c=1$. We also observe that the dynamic kink with 
$c > 3$ can show instability. This is not inconsistent with our 
earlier analysis of the static kink -- only the static kink is stable 
for $c \ge 3$ and there is no reason why rigidly moving dynamic kinks 
should be stable when $c \ge 3$.

\section{Field theoretic problem with external potential}
\label{fullproblem}

A drawback of the linearized analysis is that it keeps the
background ``rigid'' -- the profile function is always a
hyperbolic tangent -- while an oscillating kink should also
experience some profile fluctuations and be more like a
``fluid''. Hence in our second analysis of the problem we
treat the full field theory, rather than perform a perturbative
analysis about a chosen field background. To obtain a dynamic 
embedded kink, we have introduced a background potential in
which the embedded kink can oscillate. The Lagrangian density now
is given by eq. (\ref{Lagrangian}) together with an additional
``external'' term in the potential:
\begin{equation}
V_{ext}(\phi ) = {1\over 2} U(x) (\phi_1^2 - 1 )^2
\label{Vphi}
\end{equation}
where $U(x)$ is a potential well in which the embedded kink oscillates. 
The external potential can possibly originate due to the expectation 
value of other very massive fields on which backreaction is
unimportant\footnote{This could be realized physically in a
system which has a phase containing only unstable kinks (call it 
the A phase) that is sandwiched between layers of another phase 
(call it the B phase) in which the kinks are stable. Then 
the B-phase stable kink could be driven to oscillate across a 
region of A-phase where it would be unstable. The potential 
$V_{ext}$ contains both the driving potential ($U(x)$) and 
also the spatial variation of the phases because of the field 
dependence in it.}.  
A simple example for the external potential is 
$U(x) = d \, x^2/2$  ($d > 0$). Then $V_{ext}$ provides a force 
between the position of the embedded kink (inside which 
$\phi_1 \sim 0$) and the origin of the coordinate system 
($x=0$). Hence an embedded kink that is initially displaced from 
$x=0$, oscillates about $x=0$ and the frequency of oscillation is 
determined by the parameter $d$. In our numerical analysis, we have 
chosen
\begin{equation}
U(x) = {d\over 2} \alpha^2 (1-e^{-(x/\alpha)^2})
\label{U(x)}
\end{equation}
and we have chosen $\alpha = 15$, $d=0.1$. 
Then $U(x)$ has the $x^2$ form for small $x$ and changes
to a constant for $x >> \alpha$. The feature that $U(x)$
does not diverge at large $x$ has the advantage that radiation of
$\phi_1$ waves from the oscillating kink can escape to infinity.

Now the equations of motion for $\phi_1$ and $\phi_2$ are:
\begin{equation}
{\ddot \phi}_1 - {\phi}_1 '' + (\phi_1^2 + c\phi_2^2 - 1)\phi_1
     + 2 U (\phi_1^2 - 1 )\phi_1  =0
\label{phi1eq}
\end{equation}
\begin{equation}
{\ddot \phi}_2 - {\phi}_2 ''+ (c \phi_1^2 + \phi_2^2 -1)\phi_2=0 \ .
\label{phi2eq}
\end{equation}

The equations of motion are first solved numerically with the initial 
condition:
\begin{equation}
\phi_1 (t=0,x) = \tanh ({\bar X}(0,x)) \ , \ \
\phi_2 (t=0,x) = 0 \ .
\label{phi1att=0}
\end{equation}
where
\begin{equation}
{\bar X} (0,x) = \sqrt{1+2U(A)}\, X(0,x)
\label{Xbar}
\end{equation}
The factor of $1+2U(A)$ arises since, as seen in eq. (\ref{phi1eq}), 
the external potential contributes to the width of the kink.
On evolution the amplitude of oscillations is expected to decay due
to radiation. However, the decay is very slow and we can
easily track the evolution of the fields through several
tens of oscillations with the amplitude of oscillation
nearly constant. Another feature worth pointing out is that 
the angular frequency of oscillations depends on the amplitude 
of oscillation. The angular frequencies of the runs with 
large initial amplitudes are $\sim 1/A$ but at small
amplitudes the angular frequency saturates at $\sim 0.3$.

Next we ran our numerical code with a non-vanishing perturbation.
The initial condition for $\phi_1$ is still given by 
eq. (\ref{phi1att=0}) but $\phi_2$ is chosen to be given
by 
\begin{equation}
\phi_2 (t=0, x) = \epsilon \, {\rm sech} ({\bar X}(0,x))
\label{phi2t=0full}
\end{equation}
where $\epsilon$ is a very small number.
The numerical analysis is done with fixed external potential.

\begin{figure}
\scalebox{0.80}
{\includegraphics{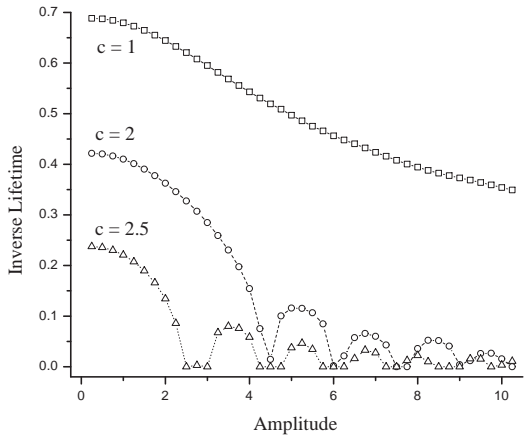}}
\caption{\label{fulltheorybands}
The decay constant $\tau^{-1}$ as a function of the initial
amplitude of oscillation in the full field theory for 
$c=$ 1 (solid line), 2 (dashed line), 2.5 (dotted line). Here 
too we observe bands of stability though none occur for $c=1$.
}
\end{figure}

As in the linearized analysis, 
a plot of $\phi_2 (t,x)$ versus time either shows exponential 
growth (superposed on oscillations), or only oscillations. This 
allows us to calculate the rate of growth of the perturbations. 
Once again we find bands of stability (see Fig. \ref{fulltheorybands}).
The new feature is that these bands occur even for rather large
amplitudes showing that dynamical (and not only ``jittering'')
embedded kinks can be stable. Note that stability bands do not
occur in the $U(1)$ model ($c=1$) just as in the linearized
analysis.

\section{Conclusions and discussion}
\label{conclusions}

Our analysis of the dynamical embedded kink has shown that
there are stability bands in field theories, similar to those 
for the Mathieu equation. This means that for certain dynamics
the embedded kink is stable to perturbations. In these cases, 
the decay of the embedded kink is determined by the rate of 
radiation and hence its lifetime can be much longer than the 
lifetime determined by considering the decay of a static embedded 
kink.

It is worth commenting on the differences between the linearized and
full field theory approaches since the results are quite different.
In the linearized case, the oscillating background is inserted
by hand and this gives stability bands occurring when the kink
has the largest acceleration. In the full field theory, an 
external potential is imposed, supposedly arising from an
external field. Now the kink oscillates naturally, but we 
observe stability bands at large amplitudes. This feature is 
explained by our earlier remark that the external potential 
influences the structure of the kink.
As the kink gets further away from the center of attraction,
it gets thinner and hence more stable. (The connection between
the width of a static kink and the instability eigenvalue
can be derived quite easily from a generalization of the
solution in eq. (\ref{phi2solution}) as given in 
Ref. \cite{MorFes53}.) Then the reason why kinks with large 
oscillation amplitudes show more stability is that they
spend more time in regions where the kink is thin and
relatively stable. This is in addition to the effect described
in the introduction: the unstable mode moves in and out of 
the kink, and can grow only while it lies within the kink.
Yet the novel feature of stability and instability bands 
must still be understood in terms of resonance between the
oscillation frequency and the phase of the unstable mode
when it enters the kink.

A natural question at this point is -- can the tension forces
that cause extended defects such as domain walls and strings
to oscillate be described by an external potential of the type 
we have used? The answer is in the negative, since the tension
forces do not cause changes in the thickness of the defects.
The only reason for the defect thickness to change is due to
Lorentz contraction. This is our reason for believing that an
extended defect will be more accurately described by the linearized
analysis of the previous section and not by the full field theory
analysis. We are currently extending our analysis to higher
dimensions with the aim of studying extended defects.

In the particle physics context, the results are most relevant
for the Z-string solutions in the standard model of the electroweak
interactions. In 1977 Nambu \cite{Nam77} had estimated the lifetime
of segments of Z-string (``dumbells'') by calculating 
the electromagnetic emission from the magnetic monopoles at the ends 
of a rotating dumbell. If, however, we have a closed loop of Z-string,
there are no magnetic monopoles that can lose energy by emitting
electromagnetic radiation and the only decay processes are by
radiation of massive Higgs and gauge bosons, and by the structural
instabilities of the Z-string discussed in Ref. \cite{JamPerVac93}.
(These may be understood in terms of W-condensation \cite{AmbOle90} 
in a Z magnetic field background \cite{Per93, Vac93, Vac95, Achetal94}.) 
If we assume that the Z-string in 3+1 dimensions also exhibits stability 
bands, it would mean that the lifetime of certain Z-string loops may be 
determined by their radiation rate \cite{EveVacVil84} and not by the 
instability growth rate of a static Z-string. 
This would yield a much longer lifetime and would open the possibility 
that Z-string loops might be produced in future accelerator experiments 
and might survive long enough to enable detection. 

Some cautionary remarks in extrapolating our results to Z-string loops 
are in order. The Z-string loop is an extended object and if it is 
to have an extended lifetime, every point on the loop should lie in a 
stability band. It may even turn out that the Z-string is like the $c=1$
embedded kink and that there are no stability bands for any dynamics.
Furthermore, the dynamics of the string is due to tension and not
due to an external potential. Hence the Z string stability
problem could be like the linearized example, in which case only
``jittering'' Z strings might be stable. The Z string loop can also 
have angular momentum and this will be an important factor in its 
evolution. (The stability of spinning string solutions has been
studied in \cite{Per94}.) Hence we feel that it is important to directly 
investigate the lifetime of Z-string loops, keeping in mind the possibility 
of stability bands.

\begin{acknowledgments} 
This work was supported by DOE grant number DEFG0295ER40898 at CWRU.
\end{acknowledgments}

\appendix

\section{}
\label{appendixA}

The following derivation was motivated by Ince's derivation of Floquet's 
Theorem \cite{Inc44}. Consider a differential equation of the form:
\begin{equation}
{\ddot \phi } + F(t,x)\phi \equiv H\phi=0 \, .
\end{equation}
If $H$ is Hermitian over an $N$ dimensional Hilbert space, we can 
expand this as:
\begin{equation}
H\phi=\sum_{i=1}^{N} \lambda_i \phi_i(t,x)=0
\end{equation}   
where the $\phi_i$ span the Hilbert space. Now, if $F(t,x)=F(t+T,x)$ 
where $T=2\pi /\omega$ in the notation of the earlier sections, 
then for each solution $\phi_i(t,x)$, there is a solution
$\phi_i(t+T, x)$ by the symmetry in $H$. Since the $\phi_i$ are complete, 
\begin{equation}
\phi_i(t+T ,x)=\sum_{j=1}^N a_{ij} \phi_j(t,x) \, .
\end{equation}   
Denoting the matrix $[a_{ij}]$ by ${\cal A}$ and the $N\times N$
identity matrix by ${\bf 1}_N$, for
each non-degenerate root $s_i$ of the characteristic equation
\begin{equation}
{\rm det} ( {\cal A} - s {\bf 1}_N ) =0
\end{equation}
we can construct a basis $\psi_i$ from the $\phi_i$ such that
\begin{equation}
\psi_i(t+T , x)=s_i\psi_i(t,x) \, .
\end{equation}
Multiplying both sides of the above result by $e^{-i\nu_i (t+T)}$, 
we see
\begin{equation}
e^{-i\nu_i (t+T)}\psi_i(t+T ,x) =
s_ie^{-i\nu_i T}\left[ e^{-i\nu t}\psi_i(t,x) \right] \, .
\end{equation}
Then it is clear that $e^{-i\nu t}\psi_i(t,x)$ is periodic in 
$t$ with period $T$ if $s_i = e^{+i\nu_i T}$.  
Therefore, solutions of the form 
\begin{equation}
\psi_i (t,x)=e^{i\nu_i t}P(t,x)
\end{equation}
exist to the equation $H\psi_i=0$ with $P(t+T, x)=P(t,x)$ and
$\nu_i$ given by $e^{+i\nu_i T}=s_i$ for non-degenerate $s_i$.


\begin{thebibliography}{}

\bibitem{MorFes53}
P. Morse and H. Feshbach, ``Methods of Theoretical Physics'',
Vol. 1, McGraw-Hill, New York (1953).

\bibitem{VacBar92}
T. Vachaspati and M. Barriola,
Phys. Rev. Lett. {\bf 69}, 1867 (1992).

\bibitem{BarVacBuc94}
M. Barriola, T. Vachaspati and M. Bucher,
Phys. Rev. {\bf D50}, 2819 (1994).

\bibitem{Vac92}
T. Vachaspati,
Phys. Rev. Lett. {\bf 68}, 1977 (1992); {\bf 69}, 216E (1992).

\bibitem{AchVac99}
A. Ach\'ucarro and T. Vachaspati,
Phys. Rep. 327, 347 (1999).

\bibitem{Raj87}
R. Rajaraman, ``Solitons and Instantons'', Chapter 2,
North Holland, Amsterdam (1987).

\bibitem{Nam77}
Y. Nambu,
Nucl. Phys. {\bf B130}, 505 (1977).

\bibitem{JamPerVac93}
M. James, L. Perivolaropoulos and T. Vachaspati,
Nucl. Phys. {\bf B395}, 534 (1993).

\bibitem{AmbOle90}
J. Ambjorn and P. Olesen,
Int. J. Mod. Phys. {\bf A5}, 4525 (1990).

\bibitem{Per93}
W.B. Perkins,
Phys. Rev. {\bf D47}, R5224 (1993).

\bibitem{Vac93}
T. Vachaspati,
Nucl. Phys. {\bf B397}, 648 (1993).

\bibitem{Vac95}
T. Vachaspati,
in Proceedings of the NATO workshop on ``Electroweak Physics
and the Early Universe'', eds. J.C. Romao and F. Friere,
Sintra, Portugal (1994), Series B: Physics Vol. 338, Plenum
Press, New York (1994).

\bibitem{Achetal94}
A. Ach\'ucarro, K. Kuijken, L. Perivolaropoulos and T. Vachaspati,
Nucl. Phys. {\bf B388}, 435 (1992).

\bibitem{EveVacVil84}
T. Vachaspati, A.E. Everett and A. Vilenkin,
Phys. Rev. {\bf D30}, 2046 (1984).

\bibitem{Inc44} 
E. Ince, 
``Ordinary Differential Equations'', Chapter 15,
Dover Publications, New York, (1944). 

\bibitem{Per94}
L. Perivolaropoulos,
Phys. Rev. {\bf D50}, 962 (1994).

\end{thebibliography}
\end{document}